# Predictive Wand: a mathematical interface design for operations with delays

Masaki Isono, Hideyoshi Yanagisawa*

*Graduate School of Engineering, The University of Tokyo, 7-3-1 Hongo, Bunkyo-ku, Tokyo 113-8656, Japan*
\* Corresponding author.
*E-mail address*: hide@mech.t.u-tokyo.ac.jp

**ABSTRACT**

Action-feedback delay during operation reduces both task performance and sense of agency (SoA). In this study, using information-theoretic free energy, we formalized a novel mathematical model for explaining the influence of delay on both task performance and SoA in continuous operations. Based on the mathematical model, we propose a novel interface design called Predictive Wand for predicting future outcomes to prevent task performance and SoA degradation resulting from response delays. Model-based simulations and operational experiments with participants confirmed that operational delay considerably reduces both task performance and SoA. Furthermore, the proposed Predictive Wand mitigates these problems. Our findings support the model-based interface design for continuous operations with delay to prevent task performance and SoA degradation.

## 1. Introduction

Advances in information technology have enabled remote operations in several applications. However, action-feedback delays typically occur in remote operations depending on the status of the network. For example, if a delay occurs in the remote control of a mobile robot, the response to operational inputs such as start/stop and direction change are delayed, which may result in an accident. A response delay diminishes the sense of agency (SoA), which refers to the perception of control over actions and their consequences (Haggard and Tsakiris, 2009). Blakemore et al. (1999) studied the effect of delay on the perception of self-produced stimuli using a self-touch paradigm and revealed that people experienced tickling when a delay occurred between voluntary action and tactile stimuli. Farrer et al. (2008) determined that delayed visual feedback caused people to perceive that they were viewing temporally displaced movements. Yang and Yanagisawa (2021) verified that delay diminishes the SoA in the same manner for both discrete and continuous operations. Studies have investigated the relationship between delays and SoA (e.g., Oishi et al., 2018; Rossetti et al., 2022; Shimada et al., 2009; Wen et al., 2019) and proven that the absence of SoA results in a person feeling less responsible for the operation (Haggard and Tsakiris, 2009; Moore, 2016; Moretto et al., 2011). Therefore, designing interfaces that prevent delays from reducing the

SoA as well as task performance is critical. Automation mitigates the problem of SoA loss (Wen et al., 2015); however, excessive automated operations can diminish SoA (Ueda et al., 2021; Zanatto et al., 2021).

In this study, we proposed a novel visual assistance interface named Predictive Wand to prevent the degradation of SoA and task performance resulting from delays in continuous operations. Because the quantitative relationship between delay and SoA depends on task settings (Wen et al., 2019), we devised an interface based on a mathematical model for considering the specifications of operation systems. First, we present the mathematical modeling of task performance and SoA in delayed continuous operations (Chapter 2), followed by a derivation of Predictive Wand (Chapter 3). We conducted model simulations on the effects of delay expectation, delay variance, and Predictive Wand on task performance and SoA and developed hypotheses for experiments based on simulation results (Chapter 4). We experimentally validated the effects of delay expectation, delay variance, and Predictive Wand on task performance and SoA (Chapter 5). Next, we detail the results (Chapter 6). Finally, we describe the conclusions (Chapters 7 and 8).

## 2. Modeling
*2.1 Delayed continuous operation model*

In this section, we formulate delayed continuous operations. We applied the comparator model proposed by Frith et al. (2000) to our delayed operation model (Fig. 1). Initially, this model was used to represent the motor control system to explain schizophrenia symptoms (Blakemore et al., 2002; Frith et al., 2000). Subsequently, Synofzik et al. (2008) applied this model to explain SoAs (see Section 2.4). Frith et al. (2000) postulated that an agent has three states (e.g., arm joint angle), namely the desired, predicted, and estimated actual states. The desired state is the future target state, and to achieve this state, inverse models generate motor commands, which are used by the forward model to generate the predicted state. Actual state transition is caused by motor commands, and the agent observes their execution as sensory feedback. Finally, the agent estimates the actual state based on the observations. The three states were then compared. The results revealed that these states did not differ when controls are normal (Blakemore et al., 2002; Frith et al., 2000). We used this model to formulate delayed operations.

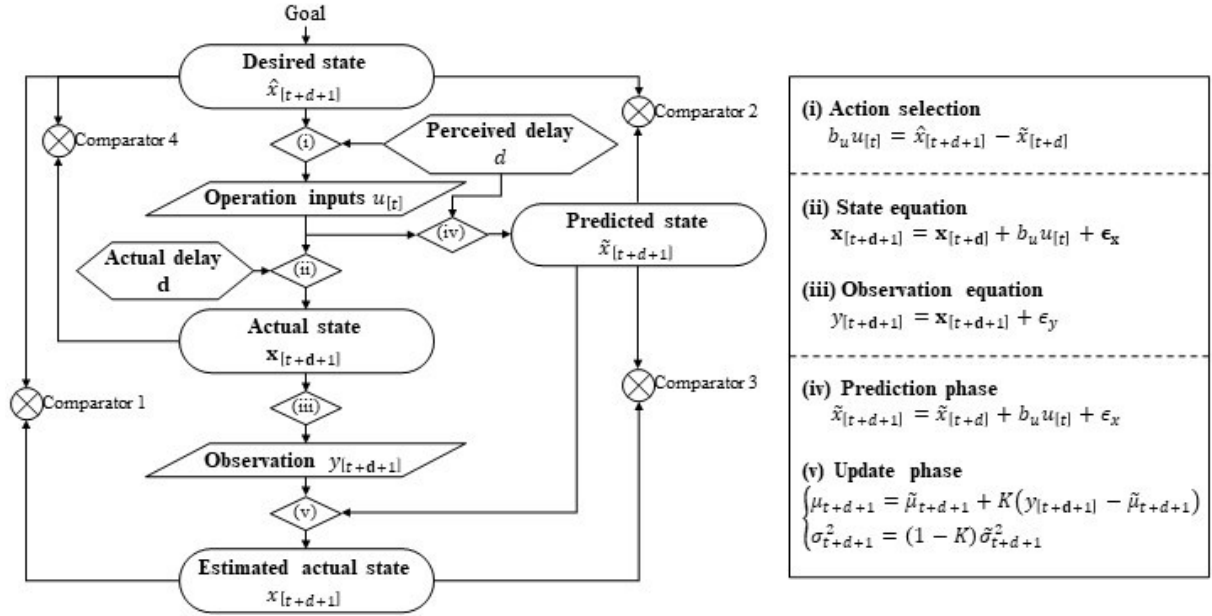

**Fig. 1.** Delayed operation model. First, the desired state is generated from the goal for operation. (i) To achieve the desired state, the operator's controllers (inverse models that convert perception to movement) work to generate operation inputs. (ii) Inputs are delayed from reaching the operation object, and state transition occurs. (iii) The operator observes state transition. (iv) The efference copy of the delayed operation inputs reach the operator's predictors (forward models that convert movement to perception), and the predicted state is generated. (v) Finally, the operator estimates the actual state from the observation and predicted state. The processes (ii) and (iii) correspond to the state-space model. Processes (iv) and (v) correspond to the Kalman filter (Kalman, 1960). The model has four states and four comparators. The comparator between the desired and estimated actual states (Comparator 1) represents the estimated operation error. The comparator between the desired and predicted states (Comparator 2) represents unpredictability (or, inexperienced operation). The comparator between the predicted and estimated actual states (Comparator 3) represents the prediction error. The comparator between the desired and the actual states (Comparator 4) represents the actual operation error. This figure is modified based on results from Frith et al. (2000); Synofzik et al. (2008).

We added actual delay to the state transition (represented by (ii) in Fig. 1) and the estimated delay to the internal models (represented by (i) and (iv) in Fig. 1). For example, the state represents the position of the robot in our model. We denote the desired, predicted, actual, and estimated actual states by $\hat{x}_{[t]}$, $\tilde{x}_{[t]} \sim \mathcal{N}(\tilde{\mu}_t, \tilde{\sigma}_t^2)$, $\mathbf{x}_{[t]}$, and $x_{[t]} \sim \mathcal{N}(\mu_t, \sigma_t^2)$, respectively. The prediction $\tilde{x}_{[t]}$ and estimation $x_{[t]}$ are random variables according to Bayesian estimation. In this paper, real-world variables are written in bold and internal model variables are written in italics. The variables are formulated based on the state-space model and Bayesian estimation.

Consider a state equation in which the operation input is reflected by a $D$-step delay as follows:

$$X_{[t+D+1]} = X_{[t+D]} + b_u u_{[t]} + \epsilon_x, \tag{1}$$

where $X$ is a stochastic state value (actual or internal), $D$ is the delay (actual or internal), $b_u$ is the constant parameter for the operation input, $u_{[t]}$ is the operation input that takes the continuous value, and $\epsilon_x \sim \mathcal{N}(0, \sigma_x^2)$ is state transition noise (actual or internal). The operator selects an operation input to achieve the desired future state $\hat{x}_{[t+d+1]}$ based on Eq. (1).

$$b_u u_{[t]} = \hat{x}_{[t+d+1]} - \tilde{x}_{[t+d]} \tag{2}$$

Because we focus on the effect of delay distribution, we assume the operator selects an optimal operation input $u_{[t]}$ from the expected values of the desired $\hat{x}_{[t+d+1]}$ and predicted $\tilde{x}_{[t+d]}$. The operation input is reflected in the actual state with a **d**-step delay that the operator observes. Here, bold $d$ denotes actual delay as follows:

$$\mathbf{x}_{[t+\mathbf{d}+1]} = \mathbf{x}_{[t+\mathbf{d}]} + b_u u_{[t]} + \boldsymbol{\epsilon}_\mathbf{x} \tag{3}$$

$$y_{[t+\mathbf{d}+1]} = \mathbf{x}_{[t+\mathbf{d}+1]} + \epsilon_y, \tag{4}$$

where $y_{[t+\mathbf{d}+1]}$ is the observation value, and $\epsilon_y \sim \mathcal{N}(0, \sigma_y^2)$ is observation noise. Eqs. (3) and (4) are based on the state-space model.

The operator predicts the future state when selecting an operational input based on Eq. (1). The italicized $d$ denotes perceived delay.

$$\tilde{x}_{[t+d+1]} = \tilde{x}_{[t+d]} + b_u u_{[t]} + \epsilon_x \tag{5}$$

Finally, the Bayesian operator estimates the actual state from observations (Eq. (4)) and prediction (Eq. (5)).

$$\begin{cases} \mu_{t+d+1} = \tilde{\mu}_{t+d+1} + \dfrac{\tilde{\sigma}_{t+d+1}^2}{\tilde{\sigma}_{t+d+1}^2 + \sigma_y^2}\left(y_{[t+\mathbf{d}+1]} - \tilde{\mu}_{t+d+1}\right) \\ \sigma_{t+d+1}^2 = \left(1 - \dfrac{\tilde{\sigma}_{t+d+1}^2}{\tilde{\sigma}_{t+d+1}^2 + \sigma_y^2}\right)\tilde{\sigma}_{t+d+1}^2 \end{cases} \tag{6}$$

Eqs. (5) and (6) are based on the Kalman filter (Kalman, 1960). Eqs. (2)–(6) represent the delayed continuous operation process.

*2.2 Discrepancy between the internal model and real-world process*

This study modeled the effect of delay distributions on performance and SoA; therefore, we formulate how delay causes a discrepancy between the internal and real-world states. See Appendix A for detailed description.

Term $\tilde{x}_{[t+d]}$ represents the prediction of future state $d$ steps ahead. The agent predicts $\tilde{x}_{[t+d]}$ based on the perceived delay $d$, so $\tilde{x}_{[t+d]}$ is directly affected by the delay. In this section, we formulate the expected value $\tilde{\mu}_{t+d}$ and variance $\tilde{\sigma}_{t+d}^2$ of $\tilde{x}_{[t+d]}$.

The agent predicts $\tilde{x}_{[t+d]}$ based on the memory of recent operation inputs. The greater the delay $d$, the more uncertain the memory is. We assume that the agent recalls past operation inputs step by step, and the recollect noise $\epsilon_u \sim \mathcal{N}(0, \sigma_u^2)$ is added to each step. Because we considered short delays of less than 1000 ms, we linearly approximated recent operational inputs. Under these assumptions, we obtain the expected value $\tilde{\mu}_{t+d}$ and variance $\tilde{\sigma}_{t+d}^2$ of $\tilde{x}_{[t+d]}$ as follows:

$$\begin{cases} \tilde{\mu}_{t+d} = \mu_t + b_u d u_{[t]} + b_u \dfrac{d(d+1)}{2} \Delta u \\ \tilde{\sigma}_{t+d}^2 = \sigma_t^2 + b_u^2 \dfrac{d(d+1)}{2} \sigma_u^2 + d\sigma_x^2, \end{cases} \quad (7)$$

where $\Delta u$ represents the expected change in operation input $u$ per time step. For example, during the operation of a mobile robot, the change in the operation per unit time corresponds to the frequency of turning and acceleration/deceleration. The upper row of Eq. (7) indicates that the expected value $\tilde{\mu}_{t+d}$ of the prediction $\tilde{x}_{[t+d]}$ is the estimated current actual position plus the sum of the linearly approximated recent operation inputs. The lower term indicates that the variance $\tilde{\sigma}_{t+d}^2$ of prediction $\tilde{x}_{[t+d]}$ is the sum of the uncertainty of the estimated current state $\sigma_t^2$, uncertainty of remembering recent operation inputs $\sigma_u^2$, and state transition noise $\sigma_x^2$. The larger the perceived delay $d$, the larger is the value of $\tilde{\sigma}_{t+d}^2$. Therefore, Eq. (7) indicates that perceived delay influences the probability distribution of the prediction of $\tilde{x}_{[t+d]}$.

*2.3 Modeling task performance*

In this section, we model task performance in delayed continuous operations. See Appendix A for detailed description.

We define task performance in relation to the operation error, which is the difference between the desired state $\hat{x}$ and the actual state $x$ (See Comparator 4 in Fig. 1). The operation error is as follows:

$$\mathbf{x}_{[t+\mathbf{d}+1]} - \hat{x}_{[t+d+1]}$$
$$= b_u \left( (\mathbf{d} - d) \left( u_{[t]} + \dfrac{\mathbf{d} + d + 1}{2} \Delta u \right) + \sum_{k=0}^{\mathbf{d}} \boldsymbol{\epsilon}_{\mathbf{u}[t-k]} - \sum_{k=1}^{d} \sum_{l=1}^{k} \epsilon_{u[t-k+l]} \right) + \sum_{k=0}^{\mathbf{d}} \boldsymbol{\epsilon}_{\mathbf{x}[t-k]} - \sum_{k=1}^{d} \epsilon_{x[t-k]} + 2\epsilon_{y_{[t]}}. \quad (8)$$

Equation (8) suggests that the discrepancy between the desired and actual states increases with the actual delay $\mathbf{d}$, and perceived delay $d$, and the discrepancy between them $\mathbf{d} - d$ increases. The operation error is calculated by sampling $\mathbf{d} \sim \mathcal{N}(\bar{\mathbf{d}}, \boldsymbol{\sigma}_{\mathbf{d}}^2)$, $d \sim \mathcal{N}(\bar{d}, \sigma_d^2)$, $\boldsymbol{\epsilon}_{\mathbf{u}} \sim \mathcal{N}(0, \boldsymbol{\sigma}_{\mathbf{u}}^2)$, $\epsilon_u \sim \mathcal{N}(0, \sigma_u^2)$, $\boldsymbol{\epsilon}_{\mathbf{x}} \sim \mathcal{N}(0, \boldsymbol{\sigma}_{\mathbf{x}}^2)$, $\epsilon_x \sim \mathcal{N}(0, \sigma_x^2)$, and $\epsilon_y \sim \mathcal{N}(0, \sigma_y^2)$. Task performance is defined as the percentage of acceptable operation errors.

$$(task\ performance) \equiv \dfrac{count(|\mathbf{x}_{[t+\mathbf{d}+1]} - \hat{x}_{[t+d+1]}| \leq E_{max})}{count(all)} \times 100\ [\%], \quad (9)$$

where "$count(*)$" denotes that "the count of samples that meet the condition of $*$," and $E_{max}$ represents the upper limit of the allowable operation error (e.g., road width). Eqs. (8) and (9) indicate that the greater the expected value of the delays, the greater the operation error $|\mathbf{x}_{[t+\mathbf{d}+1]} - \hat{x}_{[t+d+1]}|$ and the worse the task performance are. In Chapter 4, we numerically simulate task performance.

*2.4 Modeling SoA using free energy*

In this section, we model SoA for delayed continuous operations. The classical SoA model is known as the comparator model (Blakemore et al., 1998, 2000; Miall and Wolpert, 1996; Ohata et al., 2020; Wolpert et al., 1995). In the comparator model, the lack of SoA originates from prediction error, that is, the error between the predicted state and the estimated actual state (see Comparator 3 in Fig. 1). This model is a simple model in which the incongruence of the two states diminishes SoA. However, SoA varies continuously (Wen, 2019). Therefore, studies have proposed statistical (Wen et al., 2015) and mathematical models of SoA (Legaspi and Toyoizumi, 2019; Taniyama et al., 2021).

In this study, a free-energy model was adopted (Taniyama et al., 2021). Free energy is an information quantity that represents prediction errors in the brain (Friston et al., 2006). In the model, this quantity is used to formulate prediction errors in the comparator model. Bayesian estimation discussed in 2.1.1 is approximated by variationally minimizing the free energy (Buckley et al., 2017; Friston et al., 2006).

Free energy (Buckley et al., 2017; Friston et al., 2006) is a function related to information theory (Shannon, 1948). Free energy ($F$) is defined as the summation of internal energy and entropy.

$$F[y, Q] \equiv \mathbb{E}_{q(x)}[-\ln P(y, x)] - \mathbb{E}_{q(x)}[-\ln q(x)], \tag{10}$$

where $x$ and $y$ are the state and observation, respectively, as expressed in Eq. (1). Furthermore, $q(x)$ is the *recognition density* representing the agent's internal belief about $x$. $P(y, x)$, which is a joint probability distribution of $y$ and $x$, is a *generative model* representing the statistical model of the relationship between an observation and its causes. Equation (10) indicates that the free energy is the average deviation of a generative model prediction from the belief (or recognition). Free energy is a dimensionless quantity and can be used regardless of the units of $x$. Using Bayes' theorem, Eq. (10) is rearranged as follows:

$$F[y, Q] = D_{KL}[q(x) \| p(x|y)] - \ln p(y) \tag{11}$$

$$F[y, Q] \geq -\ln p(y). \tag{12}$$

The first term on the right side of Eq. (11) is the KL divergence between the recognition density and posterior distribution of $x$. The KL divergence is approximated to be zero when $q(x)$ is approximated to the posterior by variationally minimizing free energy. The minimalized free energy is the Shannon surprise, $-\ln p(y)$, representing the unpredictability of the observation $y$. The comparator model suggests that the agent lost the SoA when the observation of the action outcome was unpredicted. Therefore, Taniyama et al. (2021) proposed that SoA is inversely proportional to the minimized free energy.

$$(SoA) \propto -F = \log p(y) \tag{13}$$

The minimized free energy is expressed by following equations when a Gaussian generative model is assumed (Taniyama et al., 2021; Yanagisawa, 2016, 2021):

$$F = \frac{1}{2}\left(\frac{1}{s_p + s_l}\delta^2 + \ln 2\pi(s_p + s_l)\right), \tag{14}$$

where $\delta$, $s_p$, and $s_l$ represent the prediction error, prediction uncertainty, and system noise, respectively, and the free energy can be expressed as a function of three parameters. Taniyama et al. (2021) validated the free energy model of SoA through button-press task experiments, where the prediction error and uncertainty were controlled using operational delay and sensory modalities, respectively.

We combined the free-energy model (Eqs. (13) and (14)) and the delayed continuous operation model (Eqs. (1)–(6)). Taniyama et al. (2021) considered prediction error $\delta$ as the delay itself. However, in a continuous operation, the operator is assumed to know that a delay occurs. Therefore, the free energy associated with this state was considered. Prediction error $\delta$ is the difference between the expected values of the predicted state $\tilde{\mu}_{t+d+1}$ and the estimated actual state $\mu_{t+d+1}$ (Comparator 3 in Fig. 1), prediction uncertainty $s_p$ is the standard deviation of the predicted state $\tilde{\sigma}_{t+d+1}$, and system noise $s_l$ is the standard deviation of the observation noise $\sigma_y$. As described in Section 2.3, these variables were formulated as the functions of the expected value and variance of delay. See Appendix A for detailed formulation and assumptions.

$$\mu_{t+d+1} - \tilde{\mu}_{t+d+1}$$
$$= \frac{\tilde{\sigma}_{t+d+1}^2}{\tilde{\sigma}_{t+d+1}^2 + \sigma_y^2}\left(b_u(\mathbf{d} - d)\left(u_{[t]} + \frac{\mathbf{d} + d + 1}{2}\Delta u\right) + b_u \sum_{k=0}^{\mathbf{d}} \boldsymbol{\epsilon}_{\mathbf{u}[t-k]} + \sum_{k=0}^{\mathbf{d}} \boldsymbol{\epsilon}_{\mathbf{x}[t-k]} + \epsilon_{y[t+\mathbf{d}+1]} + \epsilon_{y[t]}\right) \tag{15}$$

$$\tilde{\sigma}_{t+d+1}^2 = \sigma_y^2 + b_u^2 \frac{d(d+1)}{2}\sigma_u^2 + (d+1)\sigma_x^2 \tag{16}$$

Eq. (15) suggests that the prediction error increases with the actual delay. Eq. (16) suggests that the prediction uncertainty increases as the perceived delay increases. The prediction error is calculated by sampling $\mathbf{d} \sim \mathcal{N}(\bar{\mathbf{d}}, \boldsymbol{\sigma}_{\mathbf{d}}^2)$, $d \sim \mathcal{N}(\bar{d}, \sigma_d^2)$, $\boldsymbol{\epsilon}_{\mathbf{u}} \sim \mathcal{N}(0, \boldsymbol{\sigma}_{\mathbf{u}}^2)$, $\epsilon_u \sim \mathcal{N}(0, \sigma_u^2)$, $\boldsymbol{\epsilon}_{\mathbf{x}} \sim \mathcal{N}(0, \boldsymbol{\sigma}_{\mathbf{x}}^2)$, $\epsilon_x \sim \mathcal{N}(0, \sigma_x^2)$, and $\epsilon_y \sim \mathcal{N}(0, \sigma_y^2)$. We define SoA as a value that increases with decreasing free energy. We mapped SoA values between 0 and 100 for comparison with experimental results.

$$(SoA) \equiv \frac{1}{\text{count}(all)} \times \sum_{sample} \frac{F_{max} - F_{sample}}{F_{max}} \times 100\ [\%] \tag{17}$$

$$F_{sample} = \frac{1}{2}\left(\frac{|\mu_{t+d+1} - \tilde{\mu}_{t+d+1}|^2}{\tilde{\sigma}_{t+d+1} + \sigma_y} + \ln 2\pi(\tilde{\sigma}_{t+d+1} + \sigma_y)\right), \tag{18}$$

where $F_{max}$ is a suitable constant for mapping SoA from 0 to 100. In Section 4, we numerically simulate SoA.

## 3. Predictive Wand

In this section, we detail the proposed visual interface named "Predictive Wand." As described in Section 2.2, the model suggests that the discrepancy between the predicted ($\tilde{x}_{[t+d]}$) and actual ($\mathbf{x}_{[t+d]}$) states at time $t + d$ causes degradation in both task performance and SoA. We hypothesize that accurate prediction $\tilde{x}_{[t+d]}$ reduces both operational and prediction errors. Based on this hypothesis, Predictive

Wand increases the accuracy by visualizing the state prediction at time $t + d$, as displayed in Fig. 2. We constructed a Predictive Wand using the actual current state $\mathbf{x}_{[t]}$, current operation input $u_{[t]}$, and expected value of the actual delay $\bar{\mathbf{d}}$.

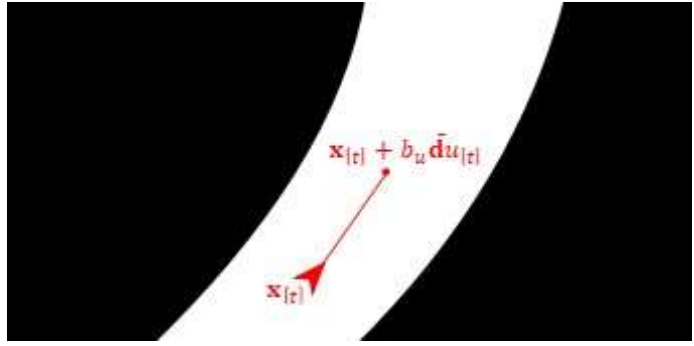

**Fig. 2.** Predictive Wand drawn as an arrow extended from the actual current state $x_{[t]}$. The length of the wand is $b_u u_{[t]} \bar{d}$. The pointed position is the observation of predicted $\tilde{x}_{[t+d]}$, and Kalman filtering is used to update prediction.

The position, $w_{[t]}$, pointed by the wand is represented by the following equation:
$$\mathbf{w}_{[t]} = \mathbf{x}_{[t]} + b_u \bar{\mathbf{d}} u_{[t]}. \tag{19}$$

Predictive Wand is drawn as an arrow extending from the current state. The length of the wall is $b_u u_{[t]} \bar{\mathbf{d}}$, so the value $w_{[t]}$ in Eq. (19) represents the position at which the wand points. The observational equation is as follows:
$$z_{[t]} = \mathbf{w}_{[t]} + \boldsymbol{\epsilon_z}, \tag{20}$$

where $\epsilon_z \sim \mathcal{N}(0, \sigma_z^2)$ is the observation noise. The agent uses a Kalman filter to predict and estimate the actual position of the wand. In this study, we assumed that the tip of the wand is a small circle and the observation noise $\sigma_z^2$ is negligibly small compared with prediction uncertainty. With this assumption, the agent predicts $\tilde{x}_{[t+d]} \sim \mathcal{N}(\tilde{\mu}_{t+d}, \tilde{\sigma}_{t+d}^2)$ as follows (Appendix B provides a detailed description):
$$\begin{cases} \tilde{\mu}_{t+d} = \mathbf{x}_{[t]} + b_u \bar{\mathbf{d}} u_{[t]} \\ \tilde{\sigma}_{t+d}^2 = \sigma_z^2 + \sigma_p^2, \end{cases} \tag{21}$$

where $\sigma_p^2$ represents the uncertainty in predicting $\tilde{x}_{[t+d]}$ from the position indicated by Predictive Wand. This result implies that the agent does not consider the pointed position as $\tilde{x}_{[t+d]}$ directly. Because $\sigma_p^2$ is not a value that can be measured directly, similar to observation noise $\sigma_y^2$, we set an appropriate value through simulations. Finally, we formulated prediction and operational errors. See Appendix B for a detailed derivation.

$$\mu_{t+d+1} - \tilde{\mu}_{t+d+1}$$
$$= \frac{\sigma_x^2 + 2\sigma_z^2 + \sigma_p^2}{\sigma_x^2 + 2\sigma_z^2 + \sigma_p^2 + \sigma_y^2} \left( b_u \left\{ (\mathbf{d} - \bar{\mathbf{d}}) u_{[t]} + \frac{\mathbf{d}(\mathbf{d} + 1)}{2} \Delta \mathbf{u} + \sum_{k=0}^{\mathbf{d}} \boldsymbol{\epsilon_{u}}_{[t-k]} \right\} + \sum_{k=0}^{\mathbf{d}} \boldsymbol{\epsilon_{x}}_{[t-k]} + \epsilon_{y_{[t+\mathbf{d}+1]}} \right) \tag{22}$$

$$\mathbf{x}_{[t+d+1]} - \hat{x}_{[t+d+1]}$$
$$= b_u \left\{ (\mathbf{d} - \bar{\mathbf{d}}) u_{[t]} + \frac{d(d+1)}{2} \Delta \mathbf{u} + \sum_{k=0}^{d} \epsilon_{\mathbf{u}[t-k]} \right\} + \sum_{k=0}^{d} \epsilon_{\mathbf{x}[t-k]} - 2\epsilon_z - \epsilon_p \qquad (23)$$

The task performance and SoA can be simulated using the definitions given in Eqs. (9) and (17), respectively. Eqs. (22) and (23) suggest that both operation and prediction errors increase when the actual delay $d$, expected value of the actual delay $\bar{d}$, and the discrepancy between them $d - \bar{\mathbf{d}}$ increase. Compared with Eqs. (8) and (15), the perceived delay $d$ does not affect operation and prediction errors of Predictive Wands. Recollect noise $\epsilon_u$ also does not affect the operation error. Thus, we hypothesize that showing Predictive Wand reduces operation and prediction errors and increases task performance and SoA. In Section 4, we numerically simulate the effects of Predictive Wand.

## 4. Model-based simulations
### 4.1 Method

We conducted a simulation assuming a specific operation. Fig. 3 presents an overview of the task.

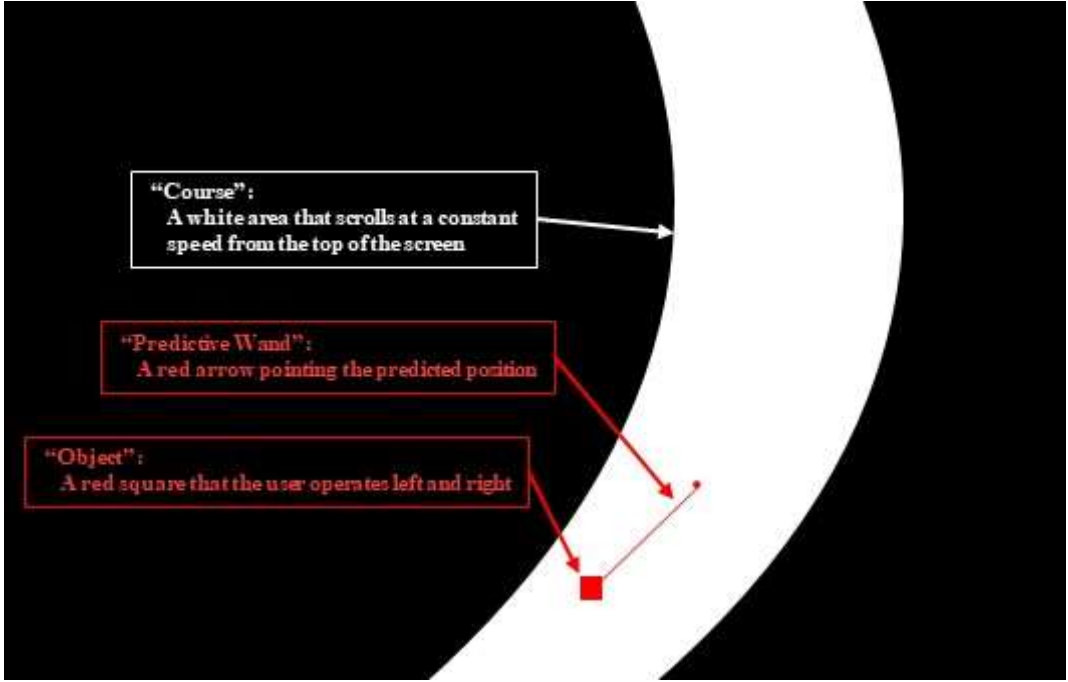

**Fig. 3**. Overview of the task for our simulation. A red object and a white course are displayed. The agent controls the object horizontally to keep it inside the course, which is scrolling vertically in a fixed speed. The task setting is identical to the experimental setting.

The contents of the task were identical to those of the experiment described in Section 5. A red square object (30 pixels on each side) and a white course (200 pixels wide) are displayed. The agent controlled the object horizontally to maintain it inside the course, which scrolled vertically at a fixed speed (200 pixel/s). A joystick was used for this operation. The deeper the tilt was, the faster was the movement of the object. The operation inputs were between −1 (maximum to the left) and 1 (maximum to the right). The max speed was 220 pixel/s. The vertical position of the object was fixed, but the scrolling speed of the course functioned as the apparent speed. The waving shape of the course was the sum of two sine waves. Predictive Wand is displayed with the condition. The wand consists of a red line segment and a red dot (five-pixel radius). The line segment length is the (apparent) velocity of the object multiplied by its mean delay. The operation delay is Gaussian distributed. The range of the mean is 200 to 1000 ms. The variance is 10 ms$^2$ (called "low variance" condition) or 1000 ms$^2$ (high variance condition).

Table 1 lists the parameters used in the simulations. The unit of time is not seconds but frames, and the rate is 100 fps.

**Table 1**

Parameters for simulation.

| Model variable | Definition | Value for simulation | Corresponding task condition |
|---|---|---|---|
| $\mathbf{\bar{d}}, \bar{d}$ | Delay expectation | 20, 40, 60, 80, 100 | Delay means |
| $\mathbf{\sigma_d^2}, \sigma_d^2$ | Delay variance | 0.1, 10 | Delay variance |
| $\mathbf{\sigma_u^2}, \sigma_u^2$ | Approximation error, Recollect noise | 0.0001 | (A suitable value) |
| $\mathbf{\sigma_x^2}, \sigma_x^2$ | State transition noise | 1.0 | (A suitable value) |
| $\sigma_y^2$ | Observation noise | 1.0 | (A suitable value) |
| $\sigma_z^2$ | Observation noise for the Wand | 1.0 | (A suitable value) |
| $\sigma_p^2$ | Uncertainty of prediction with the Wand | 400 | (A suitable value) |
| $b_u$ | Parameter for operation inputs | 2.2 | Velocity of the object |
| $u_{[t]}$ | Current input | 0.0 | A suitable value between -1 and 1 |
| $\mathbf{\Delta u}, \Delta u$ | Change of $u$ per time step | 0.005 | Course shape and scroll speed |
| $E_{max}$ | Allowable operation error | 200 | Course width |

| $F_{max}$ | Constant for mapping SoA | 500 | (A suitable value) |

To investigate changes in task performance and SoA due to delay distribution, we used the same values in the real world and the brain for the parameters in Table 1. We calculated the task performance and SoA by substituting parameters and variables in Table 1 into Eqs. (8), (9), (15–18), (22), and (23). The simulation results are presented in Section 4.2. Python was used for simulations. The simulation was performed 25 times, and the average values were plotted. A total of 5000 samples were collected for each simulation. A three-way ANOVA was conducted on both task performance and SoA with delay expectation, delay variance, and the presence of Predictive Wand.

*4.2 Results*

Fig. 4 displays the result of the simulation for task performance.

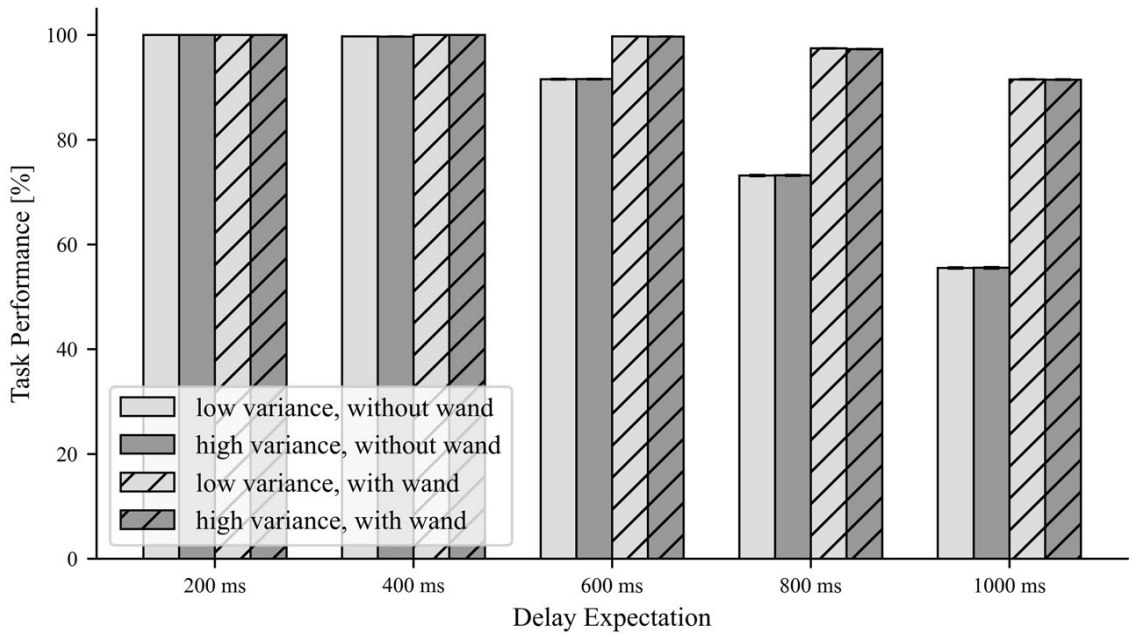

**Fig. 4**. Simulation results for task performance as a function of delay expectations for the various conditions of delay variances, and the presence of Predictive Wand. The light gray bars represent the results under low variance (i.e., 10 ms$^2$) conditions, whereas dark gray bars represent the results under high variance (i.e., 1000 ms$^2$) conditions. The hatched bars represent the results under the Predictive Wand conditions. Error bars represent standard errors.

Fig. 4 reveals that task performance declines as delay expectations increase. The statistics of three-way ANOVA reveals that the effect of delay expectation is significant ($F = 9.8 \times 10^4$, $p < 0.001$). Therefore, we hypothesize that task performance degrades with an increase in delay expectation. Fig. 4 does not suggest that the delay variance affects task performance. The main effect of the delay variance is not significant ($F = 0.069$, $p = 0.793$). We hypothesized that the delay variance does not affect task performance. Fig. 4 suggests that Predictive Wand maintained task performance under longer delay conditions (800 and 1000 ms). The effect of Predictive Wand is significant ($F = 1.8 \times 10^5$, $p < 0.001$). The interaction effect between Predictive Wand and delay expectation is also significant ($F = 4.7 \times 10^4$, $p < 0.001$). Therefore, we hypothesized that Predictive Wand reduces the rate of decline in task performance owing to an increase in delay.

Fig. 5 illustrates the result of simulation for task performance.

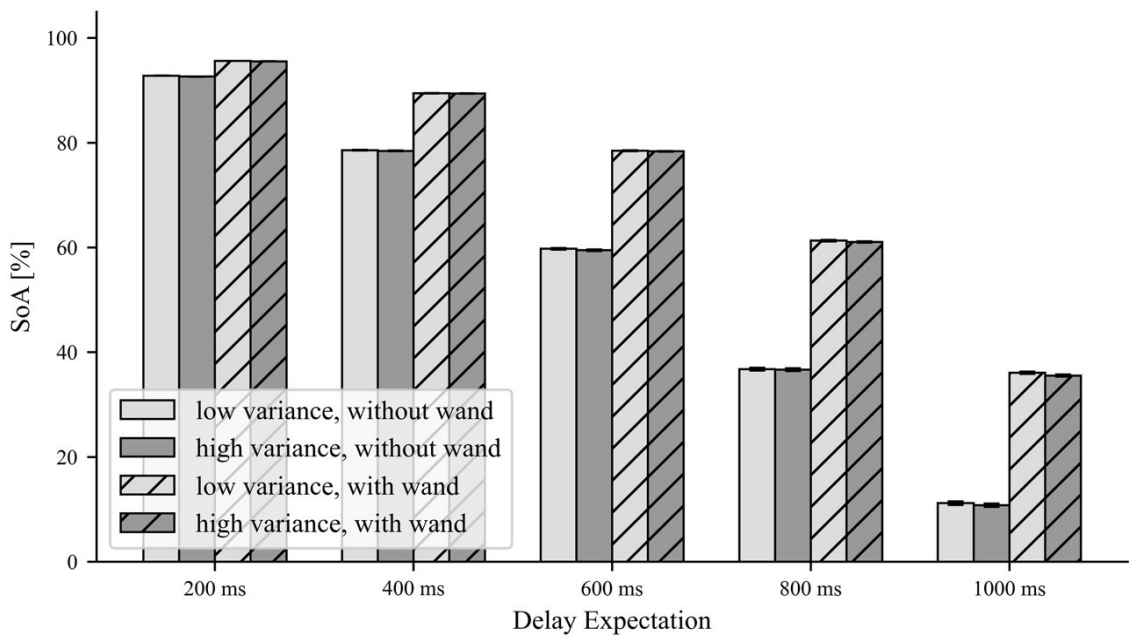

**Fig. 5.** Simulation result for SoA as a function of delay expectations for various conditions of delay variances, and presence of Predictive Wand. Light gray bars represent the results under low variance (i.e., 10 ms$^2$) conditions, whereas dark gray bars represent results under high variance (i.e., 1000 ms$^2$) conditions. Hatched bars represent the results under wand-present conditions. Error bars represent standard errors.

Fig. 5 reveals that SoA degrades with an increase in delay expectation. The statistics of three-way ANOVA revealed that the main effect of delay expectation is significant ($F = 9.7 \times 10^4$, $p < 0.001$). Therefore, we hypothesized that SoA declines with the increase in delay expectation. Fig. 5 does not suggest that the delay variance affects SoA. The main effect of delay variance is not significant ($F = 0.870$, $p = 0.351$). We hypothesized that delay variance does not affect SoA. Fig. 5 suggests that Predictive Wand prevents SoA from decreasing with an increase in delay. The main effect of Predictive Wand is significant ($F = 4.0 \times 10^4$, $p < 0.001$). The interaction effect between Predictive Wand and delay expectation is significant ($F = 2.6 \times 10^3$, $p < 0.001$). Therefore, we hypothesized that Predictive Wand reduces the rate of decline in SoA owing to a delay increase.

## 5. Experiments

*5.1 Hypotheses based on simulation results*

From the results of the simulation in Section 4.2, we proposed the following three hypotheses:

Hypothesis 1: Both task performance and SoA decrease as delay expectations increase.

Hypothesis 2: Effects of delay variance on task performance and SoA are insignificant.

Hypothesis 3: Predictive Wand reduces the rate of decline in both task performance and SoA due to an increase in delay.

The hypotheses were experimentally tested. In the experiments, participants performed operational tasks with a response delay. Delay expectation, delay variance, and presence or absence of Predictive Wand were the parameters. We measured task performance and subjective reports of SoA, and we statistically analyzed the main effects of delay expectation (for Hypothesis 1), the main effect of delay variance (for Hypothesis 2), the main effect of Predictive Wand, and the interaction effect between Predictive Wand and delay expectation (for Hypothesis 3).

*5.2 Participants*

Twenty-four university students (15 men, 6 women; mean age:22.1 ± 0.68 years) participated in the experiment. The participants had normal finger motion and sight functions. This experiment was approved by the Research Ethics Committee of The University of Tokyo Graduate School of Engineering (approval number: KE21-97). All participants consented to participate in the study.

*5.3 Procedure*

We verified the simulation results presented in Section 4.2 as hypotheses by conducting experiments with human participants. Fig. 6 displays an overview of the experiment.

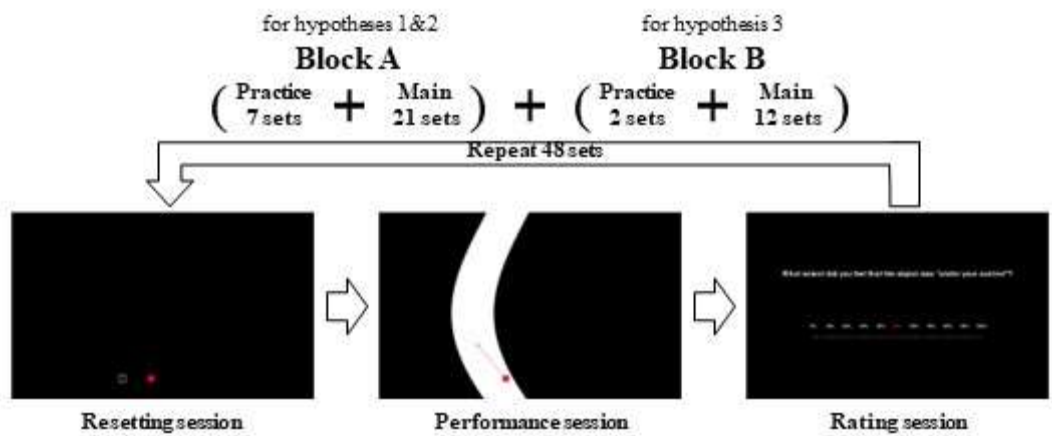

**Fig. 6.** Overview of the experiment consisting of 42 sets. The first 28 sets are called Block A and were conducted to examine the effects of delay expectation and delay variance on task performance and SoA (hypotheses 1 and 2). The remaining 14 sets are called Block B and were conducted to examine the effects

of the Predictive Wand on task performance and SoA under various delay conditions (Hypothesis 3). Each block has several practice sets at the beginning. Each set consisted of three sessions, namely resetting, performance, and rating sessions. The resetting sessions were used to reset the delay perception of participants. In the performance session, participants performed delay operation tasks, and answered subjective evaluation of SoA in the following rating session.

The experiment comprised 42 sets. The first 28 sets were called Block A. In Block A, we verified Hypotheses 1 and 2 described in Section 5.1. We examined changes in task performance and SoA with changes in delay expectations and variance. The remaining 14 sets were called Block B. In Block B, we verified Hypothesis 3 described in Section 5.1. We examined changes in task performance and SoA with and without Predictive Wand. The contents of one set of Blocks A and B were the same, with the exception for the presence of Predictive Wand as an experimental condition. The first seven sets of Block A and two sets of Block B were used for practice, but their data were not used for analysis.

Each set comprised three sessions, namely resetting, performance, and ratings. In each resetting session, a red square object and a yellow frame were displayed. Participants used a joystick to control the object horizontally. The deeper the tilt of the joystick was, the faster the object moved (maximum speed:220 pixel/s). The vertical position of the object was fixed (350 pixels from the lower edge of the monitor). Participants were instructed to control the object in the frame. When the object was in the frame, the frame moved to the other side. No delay was observed in the operation during resetting sessions. The participants repeated this task three times, and ended the session. The sessions were reset to reset the participants' perceptions of delay.

In each performance session, a red square object and a white course were displayed. Participants controlled the object as they did during reset sessions. Participants were instructed to control the object to maintain it inside the course, which was scrolled vertically at a fixed speed (200 pixel/s). The waving shape of the course was the sum of two sine waves, and three types of courses were prepared. The course width was fixed (200 pixels) from the beginning of the course to the goal. The task contents were identical to those of the simulations conducted in Section 4.

In Block A, seven conditions of delay between the tilting joystick and movement of the object were prepared: six combinations of three types of delay expectation (200, 400, or 800 ms), two types of delay variance (10 or 1000 ms$^2$), or a nondelayed condition. In Block B, 12 conditions were prepared: combinations of three types of delay expectation (200, 400, or 800 ms), two types of delay variance (10 or 1000 ms$^2$), and two types of the presence of Predictive Wand (present or absent). The vertical size of Predictive Wand was fixed ( scrolling speed of the course × delay expectation), whereas the horizontal size was variable (present input value was speed of the object × delay expectation). Each performance session lasted for 45 s.

Participants answered two questions in each rating session. The first question asked, "To what extent did you feel that the object was 'under your control'?" (0%–100 %), and the second question queried,

"To what extent did you feel that you could operate 'as you desired'?" (0%–100 %). We evaluated SoA caused by the prediction error (Comparator 3 in Fig. 1), which corresponded to the simulated SoA. The second question evaluated the desirability of operation (Comparator 1 in Fig. 1) and used to distinguish it from the SoA that we wanted to verify.

Before the experiments, the participants freely controlled the object for 30 s and performed a practice session. Experiments were conducted using the following devices: JC-U4013SBK (ELECOM) for the controller and XB323QKNVbmiiphuzx (Acer) for the monitor. All experimental conditions were counterbalanced among all participants.

*5.4 Data analysis*

The scores of task performance were calculated by the following equation:

$$(task\ performance) = \frac{t_{in}}{T_{total}} \times 100\ [\%],$$

where $T_{total}$ is the total time of one task (45 s), and $t_{in}$ is the total time that the object is inside the course in each task. Subjective reports of SoA were measured using the first questionnaires in the two rating sessions. We conducted a two-way ANOVA on both the task performance score and the subjective reports of SoA with delay expectation and delay variance for Block A. For Block B, we conducted three-way ANOVA with delay expectation, delay variance, and presence of a Predictive Wand. Scores from the nondelayed condition were excluded from analysis.

**6. Experimental results**

*6.1 Effect of the delay expectation and the delay variance (Block A)*

Fig. 7 displays the effects of delay expectation and delay variance on task performance. The results indicated that task performance declined as delay expectations increased. The main effect of delay expectations is significant ($F = 32.9$, $p < 0.001$). These results suggest that task performance declines significantly as delay expectations increase, thus supporting Hypothesis 1. The main effect of the delay variance is not significant ($F = 0.101$, $p = 0.904$). These results support Hypothesis 2, indicating that delay variance does not affect task performance.

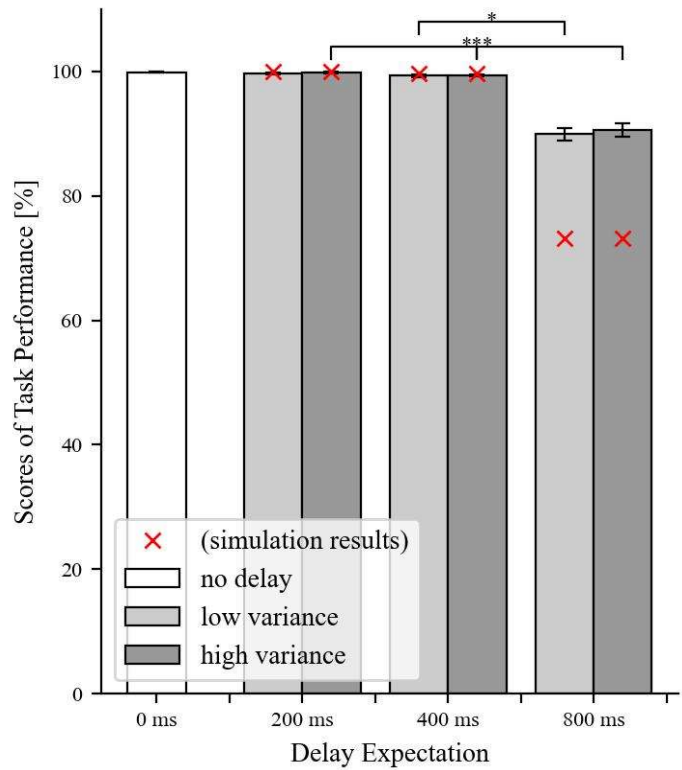

**Fig. 7.** Scores of task performance for various combinations of delay expectation and delay variance. Light gray bars represent the results under low variance (i.e., 10 ms$^2$) conditions, whereas dark gray bars represent results under high variance (i.e., 1000 ms$^2$) conditions. The leftmost bar indicates the score under nondelayed condition. Error bars represent standard errors. Red cross marks represent simulation results. Here, "*" denotes "significant ($p < 0.05$)", "***" denotes "significant ($p < 0.001$)".

Fig. 8 details the results of the effects of delay expectation and delay variance on SoA. This result indicates that SoA declines with the increase in delay expectation. The main effect of delay expectations is significant ($F = 55.2$, $p < 0.001$). This result suggests that SoA declines significantly with the increase in delay expectation, thereby supporting Hypothesis 1. The main effect of delay variance is not significant ($F = 0.377$, $p = 0.686$). This result supports Hypothesis 2: Delay variance does not affect SoA.

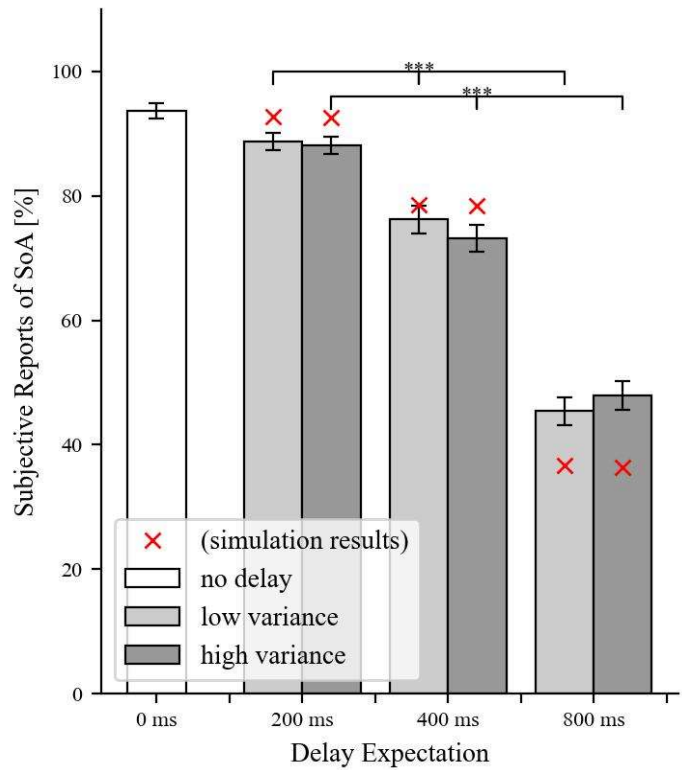

**Fig. 8.** Subjective reports of SoA for the various combinations of delay expectation and delay variance. Light gray bars represent results under low variance (i.e., 10 ms$^2$) conditions, whereas dark gray bars represent results under high variance (i.e., 1000 ms$^2$) conditions. The leftmost bar represents the score under the nondelayed condition. Error bars represent standard errors. Red cross marks represent simulation results. "***" denotes "significant ($p < 0.001$)".

*6.2 Effect of Predictive Wand (Block B)*

Fig. 9 reveals the effects of Predictive Wand and delay distributions on task performance. The result indicates that Predictive Wand maintains task performance in the 800 ms delay condition. The main effect of Predictive Wand is significant ($F = 8.039$, $p = 0.005$). The interaction effect between Predictive Wand and delay expectations is significant ($F = 4.590$, $p = 0.011$). These results suggest that Predictive Wand significantly reduces the rate of decline caused by delays in task performance, thus supporting Hypothesis 3.

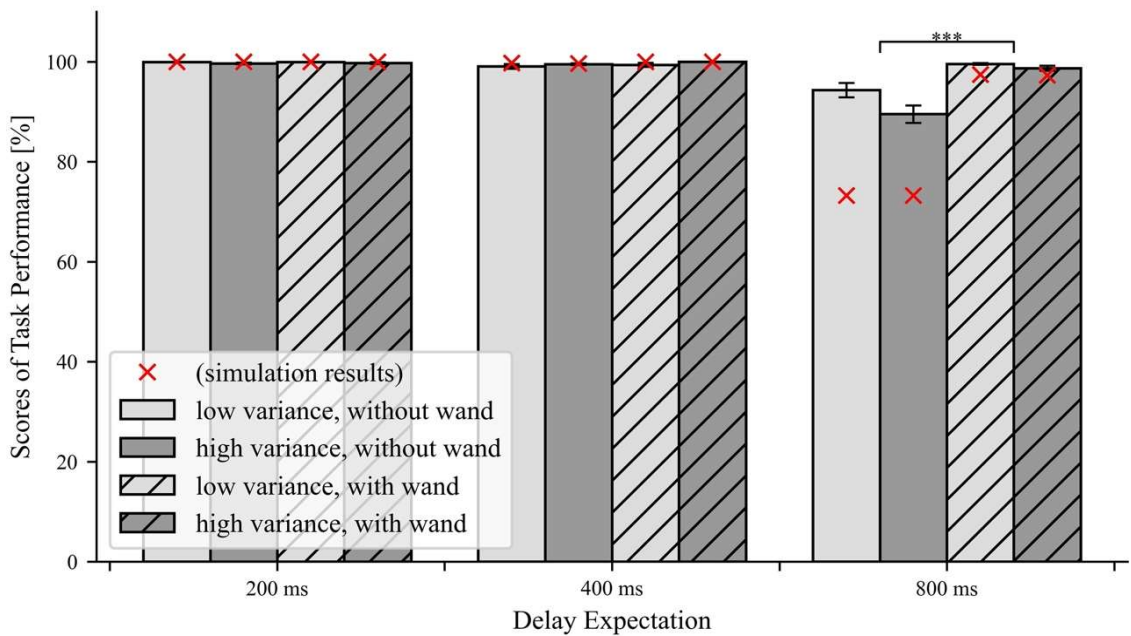

**Fig. 9.** Scores of task performance for various combinations of delay expectation, delay variance, and presence of Predictive Wand. Light gray bars represent results under low variance (i.e., 10 ms$^2$) conditions, whereas the dark gray bars represent the results under high variance (i.e., 1000 ms$^2$) conditions. Hatched bars represent the results under Predictive Wand-present conditions. Error bars represent standard errors. Red cross marks represent simulation results. "***" denote "significant ($p < 0.001$)."

Fig. 10 displays the results of effects of Predictive Wand and delay distributions on SoA. This result indicates that Predictive Wand maintains SoA in the 400 and 800 ms delay conditions. The main effect of Predictive Wand was significant ($F = 7.03$, $p = 0.0086$). The interaction effect between Predictive Wand and delay expectation was not significant ($F = 2.56$, $p = 0.0798$); however, the difference tended to increase as delay expectation increased. These results suggested that Predictive Wand significantly reduced the rate of decline owing to delays, which supported Hypothesis 3.

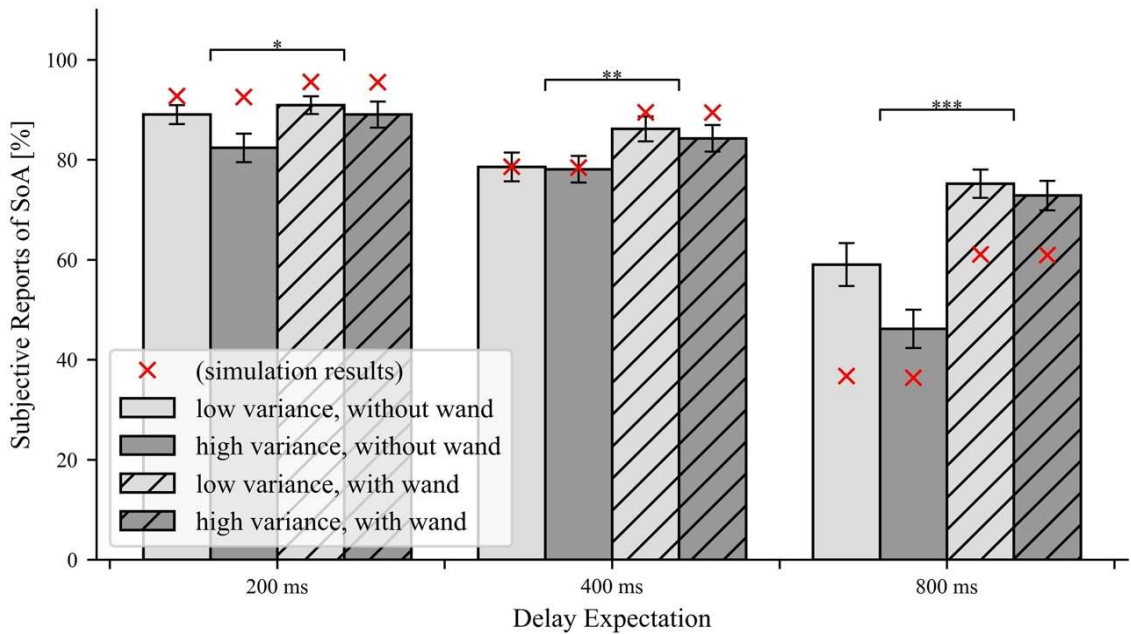

**Fig. 10.** Subjective reports of SoA for various combinations of delay expectation, delay variance, and presence of Predictive Wand. Light gray bars represent results under low variance (i.e., 10 ms$^2$) conditions, whereas dark gray bars represent results under high variance (i.e., 1000 ms$^2$) conditions. Hatched bars represent results under Predictive Wand-present conditions. Error bars represent standard errors. Red cross marks represent simulation results. "*" denotes "significant (p < 0.05)". Furthermore, "**" indicates "significant ($p < 0.01$)". "***" denotes "significant ($p < 0.001$)."

## 7. Discussions

*7.1 Effect of delay distribution on task performance and SoA*

Based on results in Section 6.1, we verify Hypotheses 1 and 2. We confirmed that task performance and SoA declined as delay expectation increased and that delay variance did not affect task performance and SoA. Figs. 7 and 8 indicate that simulation results are consistent with experimental results. For the operational tasks in this study, we consider task performance and SoA model in Chapter 2 to be appropriate. The rates of decline in both task performance and SoA were greater in simulation than in the experimental results. We assume that cause is the limitation of linear approximation of recent operation inputs $u$ (see Section 2.2). The linear approximation of $u$ becomes less valid with the increase in delay expectation. To model cases in which delay expectation is greater than approximately 800 ms, we use a method without a linear approximation of $u$.

Our simulation and experimental results on the effects of delay on task performance and SoA are consistent with those presented in other studies (e.g., Oishi et al., 2018; Rossetti et al., 2022; Shimada et al., 2009; Wen et al., 2019). This study makes it possible to measure task performance and SoA not only by experiment but also by simulation.

*7.2 Effect of Predictive Wand on task performance and SoA*

Based on the results in Section 6.2, we verify Hypothesis 3. We confirmed that Predictive Wand reduces the rate of decline by delaying task performance and SoA. Figs. 9 and 10 indicate that simulation results are consistent with experimental results. For the operational tasks in this study, we consider task performance and SoA model in Chapter 3 to be appropriate. Predictive wand indicates an alternative to automation for preventing the degradation of task performance and SoA (Wen et al., 2015; Ueda et al., 2021; Zanatto et al., 2021). Similar to the discussion in Section 7.1, we use a method without linear approximation of $u$ to model cases in which the delay expectation is greater than approximately 800 ms.

Predictive Wand is an interface calculated from the current operational input and delay expectation (see Eq. (19)). Operational inputs are not reflected in positions indicated by Predictive Wand. Therefore, when the change in the operation inputs per unit time is large, the error between the position predicted by Predictive Wand and actual position increases. The change of the operation inputs per time is represented by $\Delta u$ in our model. Fig. 11 displays a simulation of SoA when $\Delta u$ is changed from 0.005 to 0.015.

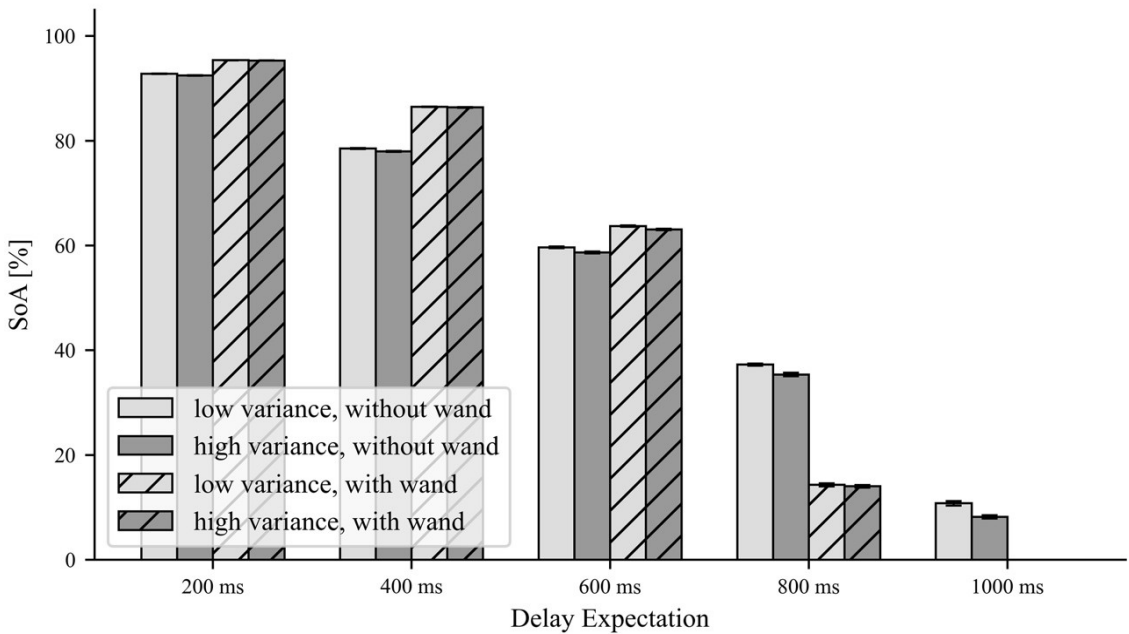

**Fig. 11.** Simulation result for SoA as a function of delay expectations for various conditions of delay variances, and presence of Predictive Wand. The difference from the results displayed in Fig. 5 is that $\Delta u$, the change of the operation inputs per time step, is changed from 0.005 to 0.015. Under 800 or larger delay expectation, SoA decreases with Predictive Wand.

Compared with Fig. 5, Fig. 11 suggests that SoA decreases with Predictive Wand under 800 or larger delay expectation. From this result, we consider that task performance and SoA rather decline with

Predictive Wand when $\Delta u$ increases, that is, when the course gets more tortuous. The relationship between $\Delta u$ and effect of Predictive Wand should be investigated in the future.

## 8. Conclusion

This study had two findings. First, a mathematical model was proposed to explain the effects of delay expectation and delay variance on task performance and SoA during continuous operation. The delayed operation model was constructed based on the state-space model and Bayesian estimation, and a prediction of the future state was formulated based on perceived delay. We derived the operation and prediction errors from the formulated prediction and modeled task performance and SoA. In both model simulations and human experiments, we confirmed the following relationships: task performance and SoA decline as delay expectation increases, and delay variance does not affect task performance and SoA.

Second, we proposed Predictive Wand, a visual interface to prevent decreases in task performance and SoA with the increase in delay. Predictive Wand is derived from Bayesian model predictions. Predictive Wand presents the prediction of future states calculated using delay expectation and current operation input. The agent predicts the future state based on Predictive Wand. This prediction diminishes operation and prediction errors. Both model simulations and experimental results confirmed that Predictive Wand reduced the rate of decline in task performance and SoA.

In conclusion, we mathematically modeled the mechanism of task performance and SoA degradation due to action-feedback delay. We verified that delay expectation, rather than delay variance, was the primary cause of task performance and SoA degradation. As delay expectation increases, task performance and SoA decrease owing to the uncertainty in predicting future states. We argue that the larger the change of the operation inputs per time, the steeper is the decrease in the task performance and SoA. Therefore, the change of the operation inputs per time must be considered for each task to estimate the task performance and SoA in the design of an operation system with delay. Predictive Wand, our novel visual interface for operation task, mitigates task performance and SoA degradation due to delay by visualizing the prediction of future states. Predictive Wand is derived from our model of the mechanism of task performance and SoA degradation due to delay. We argue that task performance and SoA degradation due to delay can be mitigated by developing interfaces that reduce the uncertainty in future state prediction based on mathematical models.

This study has several limitations. First, we used a linear approximation of recent past operation input in the mathematical model. The approximation becomes less valid with the increase in delay expectation; therefore, a method without a linear approximation of the recent past operation input was required to model cases in which delay expectation is greater than approximately 800 ms. Second, we did not consider the effect of the change in the operation inputs per time. For example, the change in operational inputs over time represents the tortuosity of courses. The simulation conducted in Section 7.2 reveal that task performance and SoA decline with Predictive Wand when the change in operation inputs per unit time is high. The relationship between the change in operation inputs over time and the effect of Predictive Wand should be studied in the future.

**Declaration of Competing Interest**

Authors declare that they have no competing financial interests or personal relationships that may have influenced this study.

**Author contributions**

**Isono Masaki:** Conceptualization, Methodology, Software, Validation, Formal analysis, Investigation, Data Curation, Writing - Original Draft, Visualization. **Yanagisawa Hideyoshi:** Conceptualization, Methodology, Validation, Resources, Writing - Review & Editing, Supervision, Project administration, Funding acquisition.

**Acknowledgments**

We thank Prof. Tamotsu Murakami and members of Design Engineering lab., Department of Mechanical Engineering, The University of Tokyo for supporting this study. We also thank Ryuichi Suzuki, Shin Shiroma, and other Sony Group researchers for the helpful suggestions through meetings.

**Funding:** This work is supported by JSPS KAKENHI grant No. 21H03528 and Sony Group corporation.

**Appendix A. Detailed formulation of task performance and SoA**

We describe the formulation of task performance and SoA in detail. The main equations are as follows: (1) state transition, (2) observation, (3) prediction, and (4) update phases (Fig. 1). The notation is the same as in Section 2.

$$\mathbf{x}_{[t+\mathbf{d}+1]} = \mathbf{x}_{[t+\mathbf{d}]} + b_u \mathbf{u}_{[t]} + \epsilon_{\mathbf{x}_{[t]}} \tag{A1}$$

$$y_{[t+\mathbf{d}+1]} = \mathbf{x}_{[t+\mathbf{d}+1]} + \epsilon_{y_{[t+\mathbf{d}+1]}} \tag{A2}$$

$$\tilde{x}_{[t+d+1]} = \tilde{x}_{[t+d]} + b_u u_{[t]} + \epsilon_{x_{[t]}} \tag{A3}$$

$$\begin{cases} \mu_{t+d+1} = \tilde{\mu}_{t+d+1} + \dfrac{\tilde{\sigma}^2_{t+d+1}}{\tilde{\sigma}^2_{t+d+1} + \sigma^2_y}\left(y_{[t+\mathbf{d}+1]} - \tilde{\mu}_{t+d+1}\right) \\ \sigma^2_{t+d+1} = \left(1 - \dfrac{\tilde{\sigma}^2_{t+d+1}}{\tilde{\sigma}^2_{t+d+1} + \sigma^2_y}\right)\tilde{\sigma}^2_{t+d+1} \end{cases} \tag{A4}$$

When the agent selects the correct operation input from the desired and predicted states, we obtain the following action-selection equation in relation to Eq. (A3).

$$b_u u_{[t]} = \hat{x}_{[t+d+1]} - \tilde{x}_{[t+d]} \tag{A5}$$

Here, $\tilde{x}_{[t+d]}$ is predicted by repeating Eq. (A3).

$$\tilde{x}_{[t+d]} = x_{[t]} + \sum_{k=1}^{d}\left(b_u u_{[t-k]} + \epsilon_{x_{[t-k]}}\right) \tag{A6}$$

The current estimated actual state $x_{[t]}$ is equal to the current predicted state $\tilde{x}_{[t]}$. Let us consider $u_{[t-k]}$ in Eq. (A6). This state represents the memory of the past input operations. Here, we assume that the agent recalls $u_{[t-k]}$ from $u_{[t-k+1]}$ as follows:

$$u_{[t-k]} = u_{[t-k+1]} + \Delta u_{[t-k+1]} + \epsilon_{u_{[t-k+1]}}, \tag{A7}$$

where $\Delta u_{[t-k+1]}$ represents the expected difference between $u_{[t-k]}$ and $u_{[t-k+1]}$. Here, $\epsilon_u \sim \mathcal{N}(0, \sigma_u^2)$ represents noise due to ambiguity of recall. Repeating Eq. (A7), we obtain the following:

$$u_{[t-k]} = u_{[t]} + \sum_{l=1}^{k} \left( \Delta u_{[t-k+l]} + \epsilon_{u_{[t-k+l]}} \right). \quad (1 \leq k \leq d) \tag{A8}$$

Substituting Eq. (A8) into Eq. (A6) gives the following expression:

$$\tilde{x}_{[t+d]} = x_{[t]} + b_u d u_{[t]} + b_u \sum_{k=1}^{d} \left\{ \sum_{l=1}^{k} \Delta u_{[t-k+l]} + \sum_{l=1}^{k} \epsilon_{u_{[t-k+l]}} \right\} + \sum_{k=1}^{d} \epsilon_{x_{[t-k]}}. \tag{A9}$$

Because we considered short delays of less than 1000 ms in this study, we assumed that $\Delta u_{[t-k]}$ is constant during $1 \leq k \leq d$. This result implies that we linearly approximate $u_{[t-k]}$.

$$\Delta u_{[t-k]} \approx \Delta u = Const. \quad (1 \leq k \leq d) \tag{A10}$$

Then, Eq. (A9) yields the following result:

$$\begin{aligned}
\tilde{x}_{[t+d]} &\approx x_{[t]} + b_u d u_{[t]} + b_u \sum_{k=1}^{d} \left\{ k \Delta u + \sum_{l=1}^{k} \epsilon_{u_{[t-k+l]}} \right\} + \sum_{k=1}^{d} \epsilon_{x_{[t-k]}} \\
&= x_{[t]} + b_u d u_{[t]} + b_u \frac{d(d+1)}{2} \Delta u + b_u \sum_{k=1}^{d} \sum_{l=1}^{k} \epsilon_{u_{[t-k+l]}} + \sum_{k=1}^{d} \epsilon_{x_{[t-k]}}.
\end{aligned} \tag{A11}$$

Because $\epsilon_u$ and $\epsilon_x$ are independent, we obtain the following equation from the linearity of normal distribution:

$$\sum_{k=1}^{d} \sum_{l=1}^{k} \epsilon_{u_{[t-k+l]}} \sim \mathcal{N}\left(0, \frac{d(d+1)}{2} \sigma_u^2\right), \quad \sum_{k=1}^{d} \epsilon_{x_{[t-k]}} \sim \mathcal{N}(0, d\sigma_x^2). \tag{A12}$$

Therefore, $\tilde{x}_{[t+d]} \sim \mathcal{N}(\tilde{\mu}_{t+d}, \tilde{\sigma}_{t+d}^2)$ can be expressed as follows:

$$\begin{cases} \tilde{\mu}_{t+d} = \mu_t + b_u d u_{[t]} + b_u \dfrac{d(d+1)}{2} \Delta u \\ \tilde{\sigma}_{t+d}^2 = \sigma_t^2 + b_u^2 \dfrac{d(d+1)}{2} \sigma_u^2 + d\sigma_x^2. \end{cases} \tag{A13}$$

Substituting Eq. (A11) into Eq. (A3), we obtain the following:

$$\tilde{x}_{[t+d+1]} = x_{[t]} + b_u(d+1) u_{[t]} + b_u \frac{d(d+1)}{2} \Delta u + b_u \sum_{k=1}^{d} \sum_{l=1}^{k} \epsilon_{u_{[t-k+l]}} + \sum_{k=0}^{d} \epsilon_{x_{[t-k]}}. \tag{A14}$$

Therefore, $\tilde{x}_{[t+d+1]} \sim \mathcal{N}(\tilde{\mu}_{t+d+1}, \tilde{\sigma}_{t+d+1}^2)$ can be expressed as follows:

$$\begin{cases} \tilde{\mu}_{t+d+1} = \mu_t + b_u(d+1) u_{[t]} + b_u \dfrac{d(d+1)}{2} \Delta u \\ \tilde{\sigma}_{t+d+1}^2 = \sigma_t^2 + b_u^2 \dfrac{d(d+1)}{2} \sigma_u^2 + (d+1)\sigma_x^2. \end{cases} \tag{A15}$$

The actual state is derived from repeating Eq. (A1) as follows:

$$\mathbf{x}_{[t+\mathbf{d}+1]} = \mathbf{x}_{[t]} + \sum_{k=0}^{\mathbf{d}} \left( b_u \mathbf{u}_{[t-k]} + \epsilon_{\mathbf{x}[t-k]} \right). \tag{A16}$$

Let us consider $\mathbf{u}_{[t-k]}$ in Eq. (A16). This model represents the past operation input. We assume that the agent recalls $u_{[t-k]}$ from $u_{[t-k+1]}$ as the following equation. We linearly approximate $\mathbf{u}_{[t-k]}$ in $1 \leq k \leq \mathbf{d}$ and denote the approximation error by $\epsilon_\mathbf{u}$.

$$\mathbf{u}_{[t-k]} \approx \mathbf{u}_{[t]} + k\Delta \mathbf{u} + \epsilon_{\mathbf{u}[t-k]}, \quad (1 \leq k \leq \mathbf{d}) \tag{A17}$$

where $\Delta \mathbf{u}$ is a gradient constant. Substituting Eq. (A17) into Eq. (A16) yields the following expression:

$$\mathbf{x}_{[t+\mathbf{d}+1]} = \mathbf{x}_{[t]} + b_u(\mathbf{d}+1)\mathbf{u}_{[t]} + b_u \frac{\mathbf{d}(\mathbf{d}+1)}{2} \Delta \mathbf{u} + b_u \sum_{k=0}^{\mathbf{d}} \epsilon_{\mathbf{u}[t-k]} + \sum_{k=0}^{\mathbf{d}} \epsilon_{\mathbf{x}[t-k]}. \tag{A18}$$

Here, we consider the differences between each state. From Eq. (A2) and the upper row of Eq. (A4), we have

$$\mu_{t+d+1} - \tilde{\mu}_{t+d+1} = \frac{\tilde{\sigma}^2_{t+d+1}}{\tilde{\sigma}^2_{t+d+1} + \sigma^2_y} \left( \mathbf{x}_{[t+\mathbf{d}+1]} - \tilde{\mu}_{t+d+1} + \epsilon_{y[t+\mathbf{d}+1]} \right). \tag{A19}$$

The left side of Eq. (A19) represents the prediction error.

From the upper row of Eq. (A15) and Eq. (A18), and Eq. (A20) when we approximate as Eq. (A21).

$$\mathbf{x}_{[t+\mathbf{d}+1]} - \tilde{\mu}_{t+d+1} = \mathbf{x}_{[t]} - \mu_t + b_u(\mathbf{d}-d)\left(u_{[t]} + \frac{d+d+1}{2}\Delta u\right) + b_u \sum_{k=0}^{\mathbf{d}} \epsilon_{\mathbf{u}[t-k]} + \sum_{k=0}^{\mathbf{d}} \epsilon_{\mathbf{x}[t-k]} \tag{A20}$$

$$\begin{cases} \mathbf{u}_{[t]} \approx u_{[t]} \\ \Delta \mathbf{u} \approx \Delta u \end{cases} \tag{A21}$$

Eq. (A21) indicates that the agent accurately perceives the current operation input and the gradient of the recent operation inputs.

From Eqs. (A5), (A11), and (A18),

$$\mathbf{x}_{[t+\mathbf{d}+1]} - \hat{x}_{[t+d+1]}$$
$$= \mathbf{x}_{[t]} - \mu_t + b_u \left( (\mathbf{d}-d)\left(u_{[t]} + \frac{d+d+1}{2}\Delta u\right) + \sum_{k=0}^{\mathbf{d}} \epsilon_{\mathbf{u}[t-k]} - \sum_{k=1}^{d}\sum_{l=1}^{k} \epsilon_{u[t-k+l]} \right) + \epsilon_t + \sum_{k=0}^{\mathbf{d}} \epsilon_{\mathbf{x}[t-k]} - \sum_{k=1}^{d} \epsilon_{x[t-k]}, \tag{A22}$$

where $\epsilon_t \sim \mathcal{N}(0, \sigma_t^2)$ is the noise due to uncertainty of the estimated actual state. The left-hand side of Eq. (A22) represents the operation error and is related to task performance.

From Eq. (A2) and the upper row of Eq. (A4), we obtain the following expression:

$$\mathbf{x}_{[t]} - \mu_t = \frac{\tilde{\sigma}_y^2}{\tilde{\sigma}_t^2 + \sigma_y^2}\left(\mathbf{x}_{[t]} - \tilde{\mu}_t\right) - \frac{\tilde{\sigma}_t^2}{\tilde{\sigma}_t^2 + \sigma_y^2}\epsilon_{y[t]}. \tag{A23}$$

This study focused on the increase in forecast uncertainty due to delays. In addition, we assumed that the agent gazes at an object on the screen. Therefore, we approximate that the observation noise is negligible compared with prediction uncertainty, that is, we have the following:

$$\sigma_y^2 \ll \tilde{\sigma}_t^2. \tag{A24}$$

Next, Eq. (A23) becomes

$$\mathbf{x}_{[t]} - \mu_t \approx \epsilon_{y_{[t]}}. \tag{A25}$$

From Eqs. (A4) and (A24), we obtain the following:

$$\sigma_t^2 = \left(1 - \frac{\tilde{\sigma}_t^2}{\tilde{\sigma}_t^2 + \sigma_y^2}\right)\tilde{\sigma}_t^2 \approx \sigma_y^2. \tag{A26}$$

From Eqs. (A19), (A20), (A22), (A25), and (A26), we obtain the following equations:

$$\mu_{t+d+1} - \tilde{\mu}_{t+d+1}$$
$$= \frac{\tilde{\sigma}_{t+d+1}^2}{\tilde{\sigma}_{t+d+1}^2 + \sigma_y^2}\left(b_u(\mathbf{d} - d)\left(u_{[t]} + \frac{d+d+1}{2}\Delta u\right) + b_u \sum_{k=0}^{\mathbf{d}} \boldsymbol{\epsilon}_{\mathbf{u}[t-k]} + \sum_{k=0}^{\mathbf{d}} \boldsymbol{\epsilon}_{\mathbf{x}[t-k]} + \epsilon_{y_{[t+d+1]}} + \epsilon_{y_{[t]}}\right) \tag{A27}$$

$$\mathbf{x}_{[t+\mathbf{d}+1]} - \hat{x}_{[t+d+1]}$$
$$= b_u\left((\mathbf{d} - d)\left(u_{[t]} + \frac{d+d+1}{2}\Delta u\right) + \sum_{k=0}^{\mathbf{d}} \boldsymbol{\epsilon}_{\mathbf{u}[t-k]} - \sum_{k=1}^{d}\sum_{l=1}^{k} \epsilon_{u[t-k+l]}\right) + \sum_{k=0}^{\mathbf{d}} \boldsymbol{\epsilon}_{\mathbf{x}[t-k]} - \sum_{k=1}^{d} \epsilon_{x[t-k]} + 2\epsilon_{y_{[t]}}. \tag{A28}$$

Now we calculate the prediction error (Eq. (A27)) and operation errors (Eq. (A28)) by sampling $\mathbf{d} \sim \mathcal{N}(\bar{\mathbf{d}}, \boldsymbol{\sigma}_\mathbf{d}^2)$, $d \sim \mathcal{N}(\bar{d}, \sigma_d^2)$, $\boldsymbol{\epsilon}_\mathbf{u} \sim \mathcal{N}(0, \boldsymbol{\sigma}_\mathbf{u}^2)$, $\epsilon_u \sim \mathcal{N}(0, \sigma_u^2)$, $\boldsymbol{\epsilon}_\mathbf{x} \sim \mathcal{N}(0, \boldsymbol{\sigma}_\mathbf{x}^2)$, $\epsilon_x \sim \mathcal{N}(0, \sigma_x^2)$, and $\epsilon_y \sim \mathcal{N}(0, \sigma_y^2)$.

Finally, we simulated task performance and SoA. In this study, we define task performance as the percentage of operational errors that are within an acceptable range.

$$(task\ performance) \equiv \frac{\text{count}(|\mathbf{x}_{[t+\mathbf{d}+1]} - \hat{x}_{[t+d+1]}| \leq E_{max})}{\text{count}(all)} \times 100\ [\%], \tag{A29}$$

where $count(*)$ indicates that "the count of samples that satisfy the condition of $*$." Furthermore, $E_{max}$ represents the upper limit of allowable operation error (e.g., road width). We calculated the SoA based on the free energy model (Taniyama et al., 2021; Yanagisawa, 2016, 2021, see Chapter 2 in detail).

$$(SoA) \equiv \frac{1}{\text{count}(all)} \times \sum_{sample} \frac{F_{max} - F_{sample}}{F_{max}} \times 100\ [\%] \tag{A30}$$

$$F_{sample} = \frac{1}{2}\left(\frac{|\mu_{t+d+1} - \tilde{\mu}_{t+d+1}|^2}{\tilde{\sigma}_{t+d+1} + \sigma_y} + \ln 2\pi(\tilde{\sigma}_{t+d+1} + \sigma_y)\right) \tag{A31}$$

Furthermore, $F_{max}$ is a suitable constant value to map SoA from 0 to 100.

**Appendix B. Detailed formulation of Predictive Wand**

We describe the detailed formulation of the effect of Predictive Wand in Chapter 3. The position indicated by Wand, $\mathbf{w}_{[t]}$ is expressed as follows:

$$\mathbf{w}_{[t]} = \mathbf{x}_{[t]} + b_u \bar{\mathbf{d}} u_{[t]} \tag{A32}$$

The following equation is its observation equation:

$$z_{[t]} = \mathbf{w}_{[t]} + \epsilon_z, \tag{A33}$$

where $z_{[t]}$ is the observation and $\epsilon_z \sim \mathcal{N}(0, \sigma_z^2)$ is the observation noise. The agent predicts $(\widetilde{w}_{[t]} \sim \mathcal{N}(\tilde{\mu}_{w_{[t]}}, \tilde{\sigma}_{w_{[t]}}^2))$ and estimates $(w_{[t]} \sim \mathcal{N}(\mu_{w_{[t]}}, \sigma_{w_{[t]}}^2))$, the actual pointed position using the Kalman Filter.

$$\widetilde{w}_{[t]} = x_{[t]} + b_u d u_{[t]} \tag{A34}$$

$$\begin{cases} \mu_{w_{[t]}} = \tilde{\mu}_{w_{[t]}} + \dfrac{\tilde{\sigma}_{w_{[t]}}^2}{\tilde{\sigma}_{w_{[t]}}^2 + \sigma_z^2}\left(z_{[t]} - \tilde{\mu}_{w_{[t]}}\right) \\ \sigma_{w_{[t]}}^2 = \left(1 - \dfrac{\tilde{\sigma}_{w_{[t]}}^2}{\tilde{\sigma}_{w_{[t]}}^2 + \sigma_z^2}\right)\tilde{\sigma}_{w_{[t]}}^2 \end{cases} \tag{A35}$$

As in Eq. (A24), we approximated the observation noise to be negligibly small compared with prediction uncertainty.

$$\sigma_z^2 \ll \tilde{\sigma}_{w_{[t]}}^2 \tag{A36}$$

Then, Eq. (A35) transforms to the following:

$$\begin{cases} \mu_{w_{[t]}} \approx z_{[t]} \\ \sigma_{w_{[t]}}^2 \approx \sigma_z^2. \end{cases} \tag{A37}$$

The agent predicts the state at time $t + d$, $\tilde{x}_{[t+d]}$ based on W and as follows:

$$\begin{cases} \tilde{\mu}_{t+d} = \mu_{w_{[t]}} \\ \tilde{\sigma}_{t+d}^2 = \sigma_{w_{[t]}}^2 + \sigma_p^2, \end{cases} \tag{A38}$$

where $\sigma_p^2$ represents the uncertainty in predicting $\tilde{x}_{[t+d]}$ from the estimated positions indicated by W and $w_{[t]}$. We considered this as a suitable constant for the simulation. From Eqs. (A3), (A32), (A33), (A37) and (A38), we obtain the following expression:

$$\begin{cases} \tilde{\mu}_{t+d+1} = \mathbf{x}_{[t]} + b_u(\bar{\mathbf{d}} + 1)u_{[t]} \\ \tilde{\sigma}_{t+d+1}^2 = \sigma_x^2 + 2\sigma_z^2 + \sigma_p^2. \end{cases} \tag{A38}$$

We now formulate the prediction and operational errors. From the upper row of Eqs. (A4), Eq. (A18), (A21), and (A38). we obtain the following expression:

$$\mu_{t+d+1} - \tilde{\mu}_{t+d+1}$$
$$= \frac{\sigma_x^2 + 2\sigma_z^2 + \sigma_p^2}{\sigma_x^2 + 2\sigma_z^2 + \sigma_p^2 + \sigma_y^2}\left(b_u\left\{(\mathbf{d} - \bar{\mathbf{d}})u_{[t]} + \frac{\mathbf{d}(\mathbf{d}+1)}{2}\Delta\mathbf{u} + \sum_{k=0}^{d}\epsilon_{\mathbf{u}[t-k]}\right\} + \sum_{k=0}^{d}\epsilon_{\mathbf{x}[t-k]} + \epsilon_{y[t+\mathbf{d}+1]}\right). \tag{A39}$$

From Eqs. (A5), (A18), (A21), (A32), (A33), (A37), and (A38), we obtain the following:

$$\mathbf{x}_{[t+\mathbf{d}+1]} - \hat{x}_{[t+d+1]}$$
$$= b_u\left\{(\mathbf{d} - \bar{\mathbf{d}})u_{[t]} + \frac{\mathbf{d}(\mathbf{d}+1)}{2}\Delta\mathbf{u} + \sum_{k=0}^{d}\epsilon_{\mathbf{u}[t-k]}\right\} + \sum_{k=0}^{d}\epsilon_{\mathbf{x}[t-k]} - 2\epsilon_z - \epsilon_p, \tag{A40}$$

where $\epsilon_p \sim \mathcal{N}(0, \sigma_p^2)$ is noise due to prediction uncertainty. Finally, we simulate task performance and SoA using the definitions given in Eqs. (A29), (A30), and (A31).